\documentclass[12pt,preprint]{aastex}
\usepackage{emulateapj5}
\usepackage{apjfonts}
\usepackage{epsfig}

\shorttitle{PKS 1510$-$089: A Head-On View}
\shortauthors{Homan et al.}
\submitted{Accepted for publication in ApJ}

\begin{document}

\title{PKS 1510$-$089: A Head-On View of a Relativistic Jet}
\author{Daniel C. Homan}
\affil{National Radio Astronomy Observatory,\altaffilmark{1}
Charlottesville, VA 22903; dhoman@nrao.edu} 
\author{John F. C. Wardle, Chi C. Cheung, David H. Roberts}
\affil{Physics Department MS057, Brandeis University, Waltham, MA
02454; wardle,cheung,roberts@brandeis.edu}
\author{and} 
\author{Joanne M. Attridge}
\affil{MIT Haystack Observatory, Westford, MA  01886; jattridge@haystack.mit.edu} 
\altaffiltext{1}{The National Radio Astronomy Observatory is a 
facility of the National Science Foundation operated under 
cooperative agreement by Associated Universities, Inc.}

\begin{abstract} 
The gamma-ray blazar PKS 1510$-$089 has a highly superluminal
milli-arcsecond jet at a position angle (PA) of $-28^\circ$ and
an arcsecond jet with an initial PA of $155^\circ$.  With a $\Delta$PA
of $177^\circ$ between the arcsecond and milli-arcsecond jets,  
PKS 1510$-$089 is perhaps the most highly misaligned radio jet ever 
observed and serves as a graphic example of projection effects
in a highly beamed relativistic jet.  Here we present the results
of observations designed to bridge the gap between the milli-arcsecond
and arcsecond scales.  We find that a previously detected
``counter-feature'' to the arcsecond jet is directly fed
by the milli-arcsecond jet. This feature is located 
$0.3\arcsec$ from the core, corresponding to a de-projected distance of
30 kiloparsecs.  The feature appears to be dominated by shocked
emission and has an almost perfectly ordered magnetic field along 
its outside edge.  We conclude that it is most likely a 
shocked bend, viewed end-on, where the jet crosses our line 
of sight to form the southern arcsecond jet.  While the bend
appears to be nearly $180^\circ$ when viewed in projection, we
estimate the intrinsic bending angle to be between $12^\circ$ 
and $24^\circ$.  The cause of the bend
is uncertain; however, we favor a scenario where the jet is bent
after it departs the galaxy, either by ram pressure due to
winds in the intracluster medium or simply by the density gradient 
in the transition to the intergalactic medium.  
\end{abstract}


\section{Introduction}
\label{s:intro}
Extragalactic radio jets can change direction for a variety of
reasons, such as deflections by massive clouds in the interstellar or 
intracluster medium, growth of hydrodynamical instabilities, or
precession at the base of the jet.  Core dominated radio sources are
oriented with their jets pointed nearly along our line of sight,
greatly exaggerating in projection any intrinsic trajectory
changes.  Small intrinsic changes can therefore appear as large
angle misalignments between the jet axes observed on parsec and 
kiloparsec scales.  Indeed, large angle misalignments have
been observed in some core dominated radio sources 
\citep[e.g. 3C\,309.1,][]{WKP86}, and the
distribution of jet misalignment angles has been extensively 
studied \citep[e.g.][]{PR88,WCU92,CM93,ASV96}.  While the distribution
is known to have a bimodal shape, with a main peak of misalignment
angles near $0^\circ$ and a 
secondary peak near $90^\circ$ \citep{PR88,CM93}, misalignment
angles larger than $120^\circ$ are quite rare.  Between two recent
studies by \citet*{TME98} of southern EGRET sources and by 
\citet*{LTP01} of the Pearson-Readhead sample, the misalignment
angles of fifty sources are compiled; just two of those sources 
have misalignment angles greater than $120^\circ$, and not one is
misaligned more than $150^\circ$.

Here we discuss the highly misaligned radio jet of PKS 1510$-$089 
($z=0.360$), a radio selected, high polarization quasar
\citep{HB93,SMA84}.  PKS 1510$-$089 has 
been detected in $\gamma$-rays by EGRET \citep{HBB99} 
and exhibits apparent jet 
motions in excess of 20 times the speed of light
\citep{H01,W02}.  
Superluminal motion is a natural consequence of a highly relativistic 
jet pointed nearly right at us \citep{R66,BK79}, leading to a 
compression of the observed time scale and magnifying intrinsic 
pattern speeds: $\beta_{app} = \beta\sin\theta/(1-\beta\cos\theta)$ where
$\beta_{app}c$ is the observed speed, $\beta c$ is the intrinsic pattern
speed, and $\theta$ is the angle between the jet axis and the line
of sight.  If traveling at the optimum angle for superluminal motion, 
$\beta=\cos\theta$, the parsec-scale jet of PKS 1510$-$089, at an
apparent speed of $20c$, lies within just three degrees of our line of sight.  

\citet*{OBC88} mapped the arcsecond scale structure of PKS 1510$-$089 
at 5, 15 and 22 GHz using the Very Large Array (VLA).  They observed a 
jet extending nine arcseconds to the south-southeast at 5 GHz and 
an oppositely directed counter-feature at just 0.3 arcseconds at the 
higher frequencies.
Three epochs of early Very Long Baseline Interferometry (VLBI)
observations at 1.7 GHz \citep[summarized in][]{BPF96} appeared to show
the milli-arcsecond scale jet directed toward the southern VLA jet.  
On the basis of these results, PKS 1510$-$089 has been considered a
well aligned source in studies of mis-alignment angles in
gamma-ray blazars \citep{TME98,HJS98,C00}.

More recent Very Long Baseline Array (VLBA) observations ranging from
2 GHz up to 43 GHz \citep{FC97,KVZC98,H01,J01,W02} show the jet extending 
up to 40 milli-arcseconds (at 2 GHz) to the 
north-northwest, directly toward the ``counter-feature'' 
at $0.3\arcsec$ observed by \citet{OBC88}.  These more recent and more
sensitive observations (as well as the deep 1.7 GHz observations presented
here) show no sign of a southern milli-arcsecond jet, suggesting 
that the early VLBI results may have simply mis-identified the core.

With a highly superluminal milli-arcsecond jet extending to 
the north-northeast,  a bright VLA feature directly in its path 
at $0.3\arcsec$, and an arcsecond VLA jet oppositely directed 
by $\simeq 180^\circ$, the connection between the jets on these scales 
is an intriguing puzzle. Here we report the results of observations 
designed to fill the gap in resolution between the previous VLBI and
VLA observations of this source.  The observations are described in
\S{\ref{s:obs}}, and in \S{\ref{s:dis}} we suggest and analyze two general 
models to explain the jet trajectory of this highly superluminal
blazar.  Throughout this paper we assume a cosmology with $H_0 = 70$ 
km s$^{-1}$ Mpc$^{-1}$ , $\Omega_m = 0.3$, and $\Omega_\Lambda = 0.7$, and
we choose a spectral index convention: $S_\nu\propto\nu^{+\alpha}$. 

\section{Observations and Results}
\label{s:obs}
On August 11, 2001 we made deep VLBI observations of PKS 1510$-$089 using 
the National Radio Astronomy Observatory's (NRAO)
VLBA plus a single VLA antenna (VLBA$+$Y1) at 1.7 and 5.0 GHz.  The 
observations were taken in dual-polarization mode so that we
could study the linear polarization of the jet as a probe of the
underlying magnetic field structure.  The data were correlated on
the VLBA correlator and were processed with
NRAO's Astronomical Imaging Processing System, AIPS, \citep{BG94,G88}
and the Caltech DIFMAP package \citep{SPT94,SPT95} using standard
techniques for VLBI polarization observations 
\citep[e.g.][]{C93,RWB94}.  The feed leakage terms were corrected using the
strong, unpolarized source OQ208.

\begin{figure*}
\figurenum{1}
\begin{center}
\epsfig{file=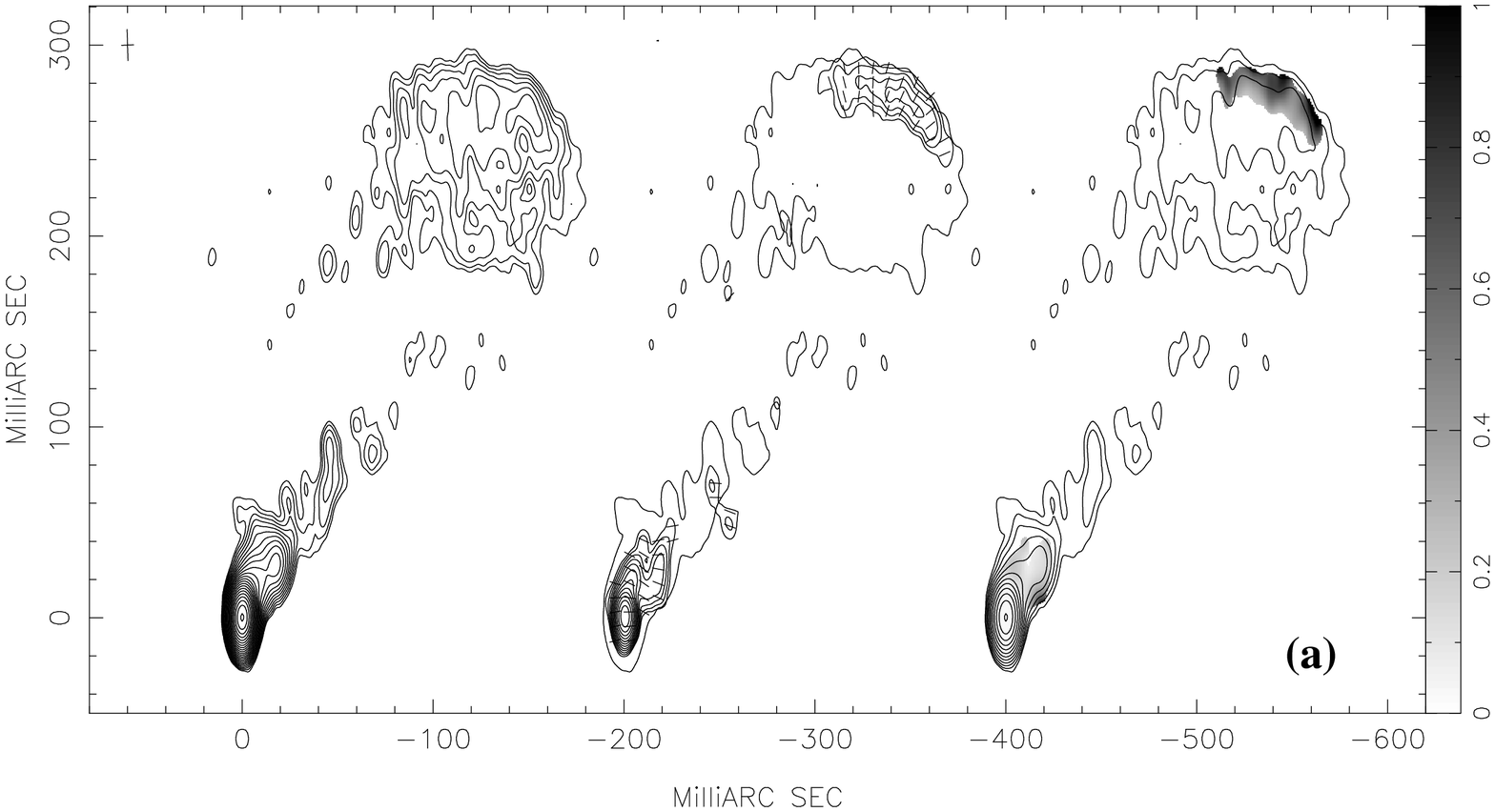,width=6.5in,angle=0}\\
\epsfig{file=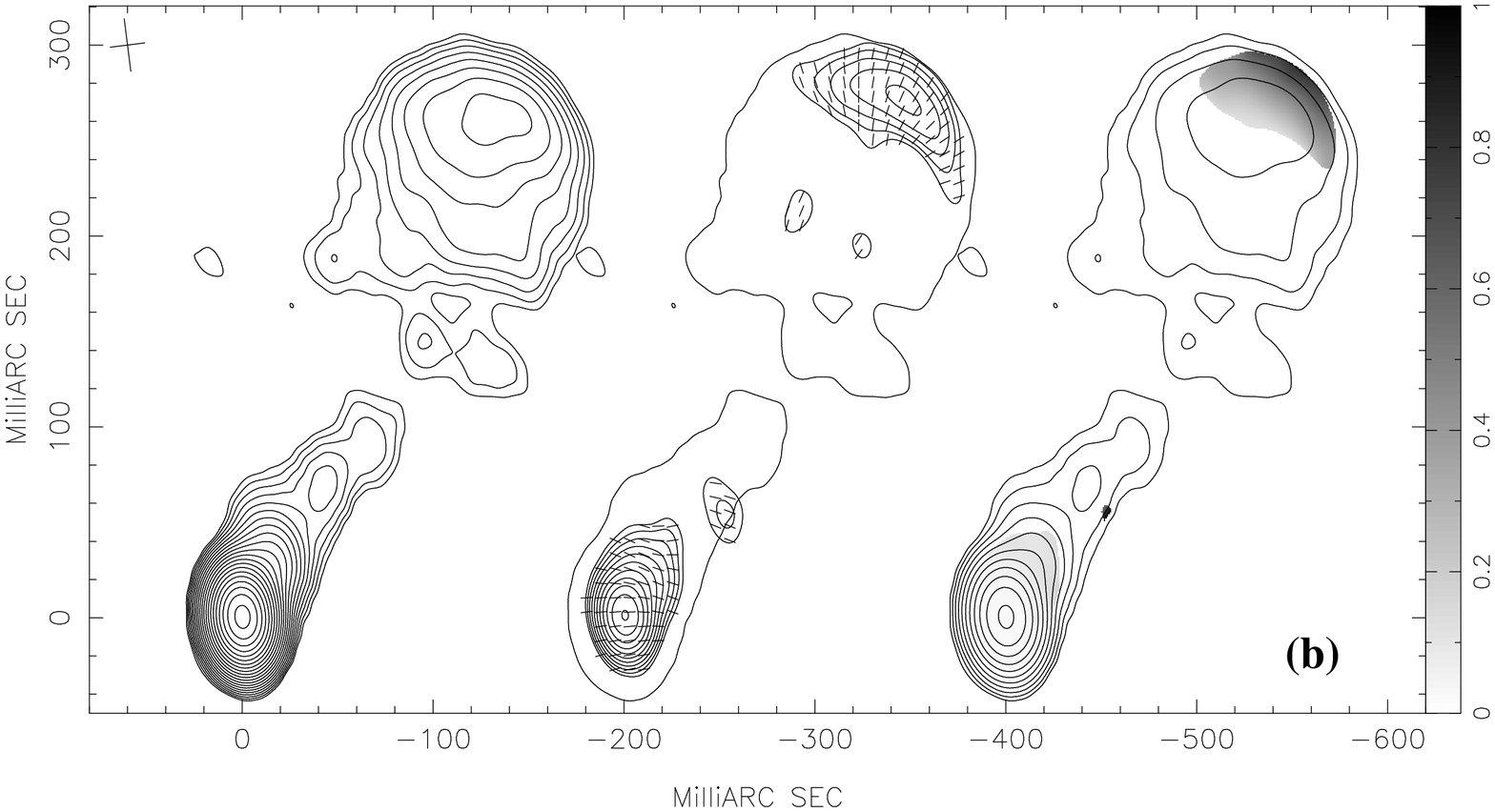,width=6.5in,angle=0}
\end{center}
\figcaption[f1a.eps,f1b.eps]{\label{f:1510-1}
Naturally weighted (above, panel (a)) and tapered (below, panel (b)) VLBA$+$Y1
images of PKS 1510$-$089 at 1.7 GHz.  The left image is total 
intensity contours ($\times\sqrt{2}$ steps), the center image is 
polarized intensity contours ($\times\sqrt{2}$ steps) with the electric
vector directions indicated by tick-marks, and the right image is
total intensity contours ($\times 2$ steps) superimposed with 
grayscale fractional polarization.  The lowest contour levels 
are $0.5$ mJy/beam in panel (a) and $1.0$ mJy/beam in panel (b).  
The FWHM sizes of the elliptical Gaussian restoring beams are 
indicated by the stick figures in the upper left hand corners of 
the panels. 
}
\end{figure*}

\begin{figure*}
\figurenum{2}
\begin{center}
\epsfig{file=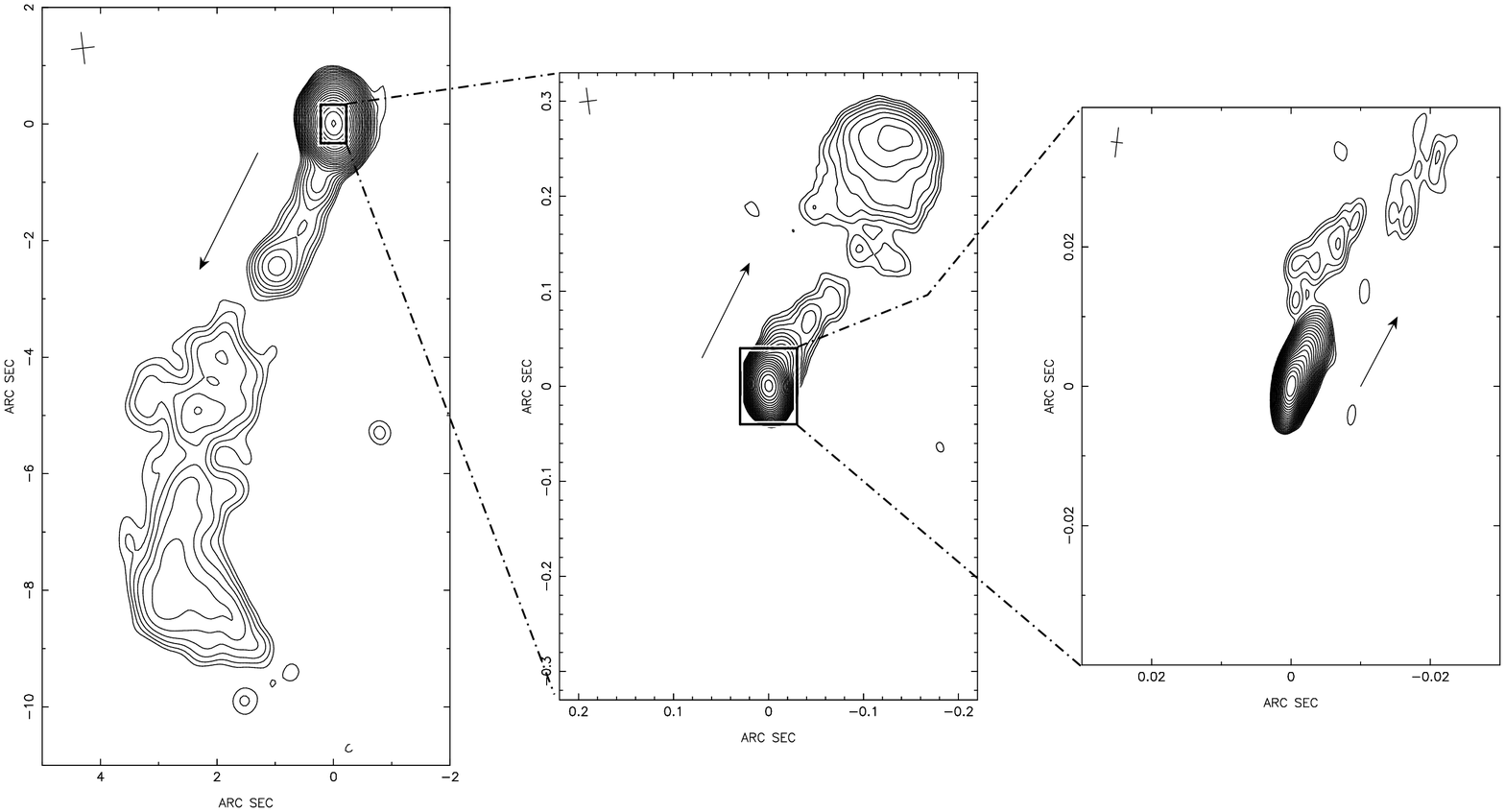,width=7in,angle=0}\\
\end{center}
\figcaption[f2.eps]{\label{f:1510-2}
The jet of PKS 1510$-$089 from arcsecond to milli-arcsecond scales. 
The left panel
is a re-processed 5 GHz VLA image from data originally published 
by \citet{OBC88}.  The center panel is the tapered version of our
1.7 GHz VLBA$+$Y1 image.  The right panel is a naturally weighted
image from our 5 GHz VLBA$+$Y1 data.  Contours levels increase in
steps of $\times\sqrt{2}$ from base levels of $0.5$ mJy/beam for the
VLA image (left panel) and $1.0$ mJy/beam for the VLBA$+$Y1 images
(center and right panels).  The FWHM dimensions of the elliptical 
restoring beams are plotted as crosses in the upper left-hand corner of 
each figure.
The arrow in each panel indicates the jet direction on that scale.
}
\end{figure*}

Figure \ref{f:1510-1} displays our naturally weighted and tapered
images of PKS 1510$-$089 at 1.7 GHz.  The tapered image was made
using a Gaussian weight taper in the (u,v)-plane with a weight
factor of 0.3 at a radius of 10 M$\lambda$.  The images
show a strong core with a jet extending to the north-northeast for
approximately 150 milli-arcseconds (mas) before fading.  The strong 
``counter-feature'' seen by \citet{OBC88} is a prominent feature 
in both the naturally weighted and tapered maps.  We do not present
polarization maps of our 5 GHz data here, although an
intensity map of the milli-arcsecond jet is included as part of figure 
\ref{f:1510-2}.  We note that, with three times the resolution of 
the 1.7 GHz images, the 5 GHz images resolve out most of the 
extended emission.  We do make use of the 5 GHz data to measure 
the spectral index of the milli-arcsecond scale jet.  

Figure \ref{f:1510-2} is a multi-frame figure showing the jet from 
arcsecond to milli-arcsecond scales.   The first frame (left) is a 
reprocessed version of the 5 GHz VLA image obtained by \citet{OBC88}.  
The second frame (center) is our tapered 1.7 GHz VLBA image with a 
wider field of view to show that there is no hint of jet emission to 
the south.  The third frame (right) is our naturally weighted 5 GHz
VLBA image of the milli-arcsecond jet.

\subsection{The Feature at $0.3\arcsec$}

At first glance the feature appears to be almost circular in shape, 
approximately 100 mas in diameter ($\simeq 500$
parsecs at $z=0.360$).  The interior of the feature 
has a sloping brightness profile that increases toward the outside
edge. The outside edge itself is defined by a sharp brightness 
gradient and is distinctly curved, almost semi-circular in shape. 
The edge toward the core is less bright and is not clearly defined.

The feature is strongly polarized in a narrowly confined region 
along part of its outside edge.  Here the fractional
polarization climbs well above 50\%, and it approaches the theoretical
maximum for synchrotron radiation ($71$\% for $\alpha=-0.6$).  Indeed,
at the very outside edge of the feature, the theoretical maximum
appears to be exceeded in places; however, polarization and intensity 
levels have sharp gradients there, and uncertainty in the
resulting fractional polarization can be large.  Based on the noise levels
in our tapered maps, we can say that the polarization at the very
outside edge exceeds 65\% at the 95\% confidence level.
The magnetic field must be almost perfectly ordered (as viewed in
projection) at this point to explain such high levels of polarization.  
This provides an important constraint on models for this feature.

The polarization vectors in the map have been corrected 
for the integrated rotation measure of $-15\pm1$ rad/m$^2$ \citep{SKB81} 
which rotates the electric vectors by $28^\circ$ at this frequency.  
The resulting polarization angles in this feature agree well with 
those measured by \citet{OBC88} at 15 GHz.  The intrinsic 
polarization vectors are very nearly perpendicular to the curved outside
edge of the feature.  The projected magnetic field ($90^\circ$ to
the polarization vectors) is therefore parallel to and curving around
the outside edge of the feature. 

\citet{OBC88} measure an integrated spectral index for the $0.3\arcsec$
feature of $\alpha \sim -0.6$ between the closely spaced frequencies
of 15 and 22 GHz, and they note that its integrated flux is $26$
mJy at $22$ GHz.  Comparing to our integrated flux of $0.13$ Jy at 1.7
GHz, we find the spectral index to indeed be $\alpha = -0.6\pm0.1$ over more 
than a decade in frequency.  At $-0.6$, the spectral index of
the feature is less steep than the milli-arcsecond scale
jet for which \citet{H02} find $\alpha = -0.9\pm0.1$ from multi-epoch VLBA
observations at 15 and 22 GHz.  Fitting simple models to our 1.7 and 
5 GHz data, we find a spectral index further out ($R\sim20-30$ mas) in the 
milli-arcsecond jet of $\alpha = -0.8\pm0.1$.

\section{Discussion}
\label{s:dis}
From figure \ref{f:1510-1}, there seems to be little doubt that 
the $0.3\arcsec$ feature is fed directly by the milli-arcsecond jet.  
In this section we explore the nature of this feature
and its relation to the milli-arcsecond and arcsecond scale jets.

\subsection{The Milli-Arcsecond Scale Jet}
\label{s:milli}

Due to its very high proper motion, we know the axis of the
milli-arcsecond jet makes a small angle with our line of 
sight. For the purposes of this discussion we take this alignment 
angle, $\theta_{VLBA}$, to be $3^\circ$, which is the optimum 
angle for the superluminal motion of 20c observed by \citet{H01}. 

The motion in the milli-arcsecond jet appears to be straight
along a structural position angle of $-28^\circ$ \citep{H01}.  This
motion points directly at the apex of the $0.3\arcsec$ feature which also
has a structural position angle of $-28^\circ$.  However, the path of the jet
from a couple milli-arcseconds to $0.3\arcsec$ is not entirely
straight, and we note that the third panel in figure \ref{f:1510-2} 
shows the jet ridge-line to wiggle in the plane of the sky after the 
first 10 mas.  The jet
ridge-line is defined by the bright parts of the jet and small wiggles 
in it will be greatly magnified by the projection effects of a 
$3^\circ$ viewing angle.  The larger scale images in figure
\ref{f:1510-1} show the ridge-line of the jet to be 
essentially straight but slightly south of the $-28^\circ$ position 
angle of the $0.3\arcsec$ feature, perhaps indicating that the jet 
feeds the southwest edge of the $0.3\arcsec$ feature.  However, it is
important to remember that the ridge-line may not trace the main 
flow, and the milli-arcsecond jet could be quite broad (and largely 
too dim to be seen), possibly filling the entire angle subtended 
by the $0.3\arcsec$ feature.  

While the wiggles of the jet ridge-line likely
represent very small bends magnified by projection, we have no
information about whether the jet bends toward or away 
from the line of sight by a significant amount.  In the absence of
better information, we will make the simple assumption that the jet 
path from milli-arcsecond scales to $0.3\arcsec$ is essentially
straight and that the position 
of the $0.3\arcsec$ feature is projected by the same $3^\circ$ angle 
as the proper motion.  This would place the
$0.3\arcsec$ feature at a de-projected distance of $30$ 
kiloparsecs (kpc) from the nucleus.  Given the $500$ pc transverse
size of the feature, the intrinsic full-opening angle of the 
milli-arcsecond jet must be $\lesssim 1^\circ$.

\subsection{The Arcsecond Scale Jet}

The bright arcsecond jet is $\simeq 9\arcsec$ long at
a structural position angle of $160^\circ$.   The inner part 
of the arcsecond jet is at a position angle of $\simeq 155^\circ$ 
\citep{OBC88}, placing it very nearly 
$180^\circ$ from the direction of the milli-arcsecond jet.  

It is interesting that the inner part of the arcsecond jet 
($1\arcsec \rightarrow 3\arcsec$), although apparently straight, does not 
form a line that extrapolates directly back to the core.  This
suggests that the
arcsecond jet has been bent somewhere from its origin to the
point where it is visible in the VLA map.  

Another interesting feature of the VLA map is that there is no
clear sign of the counter-lobe.  The counter-jet, of course, is highly
beamed away from us; however, we might expect to see 
the unbeamed emission from the counter-lobe.  In one of the models
presented below, the approaching jet bends directly across our line
of sight, forming the apparent $180^\circ$ misalignment between
the milli-arcsecond and arcsecond jets.  In this bent jet model, the
counter-lobe would lie directly behind the approaching arcsecond 
jet, and we perhaps see evidence for this at radii 
$4\arcsec\rightarrow6\arcsec$
in the 5 GHz VLA image from figure \ref{f:1510-2}.  There is
significant excess, fluffy emission in this region which does not 
appear to be directly associated with the main flow of the jet.  This
may be emission from the hidden counter-lobe which has been eclipsed
by the bent approaching jet.

\subsection{The Feature at $0.3\arcsec$}

The feature appears to be dominated by shocked emission.  It has
approximately one hundred times the surface brightness a straight
continuation of the jet would
have at this radius\footnote{We estimated the jet brightness at
$0.3\arcsec$ by taking various 1-D slices through the images and
extrapolating the power-law decay of jet brightness with radius.}.  
Doppler boosting can give at most a factor of six in brightness if the 
jet has turned directly toward us at this 
point.  If the jet has decelerated much from the $\gamma\simeq20$ flow
at its base, any gain from Doppler boosting would be much less.  
Thus shock-enhanced emission and/or very long path lengths are 
necessary to explain the brightness.  The feature also has a spectral 
index which is less steep than observed in the milli-arcsecond jet, 
suggesting a locally re-accelerated particle population.  The very high
fractional polarization of the feature, oriented transverse to the
jet direction, is also consistent with a shock, although we 
note that one possible model (described below) would allow the
polarization to be produced through shear.

Two natural models for the feature are

\begin{itemize}
\item {\em A terminal bow shock in the milli-arcsecond jet.} 
In this
model the arcsecond jet is an older jet and is not related to the
milli-arcsecond jet.  The jet direction may have 
changed either in a discrete way due to an event that profoundly 
changed the black hole/accretion disk system or in a more continuous 
fashion due to some kind of precession.  Any precession model would 
have to explain the existence of the apparent ``preferred'' jet 
direction of the arcsecond jet.
\item {\em A shocked bend in the jet.}  Here the jet bends almost
directly across our line of sight, and the arcsecond jet is
simply the continuation of the jet after the bend.
\end{itemize}

The very high levels of fractional polarization would seem to
rule out a simple bow shock model.  Given the close alignment between
the jet and the line of sight, we would expect to view the bow
shock nearly end-on, so the projected magnetic field would
appear almost completely tangled, giving little, if any, 
net polarization. For the very high levels of
fractional polarization that we see, we would have to view the bow shock 
from the side \citep[e.g.,][]{L80}.  Relativistic aberration which
can change viewing angles considerably for highly relativistic flow is
of little help here because the bow shock advance speed is likely be 
quite slow.

\begin{figure*}
\figurenum{3}
\begin{center}
\epsfig{file=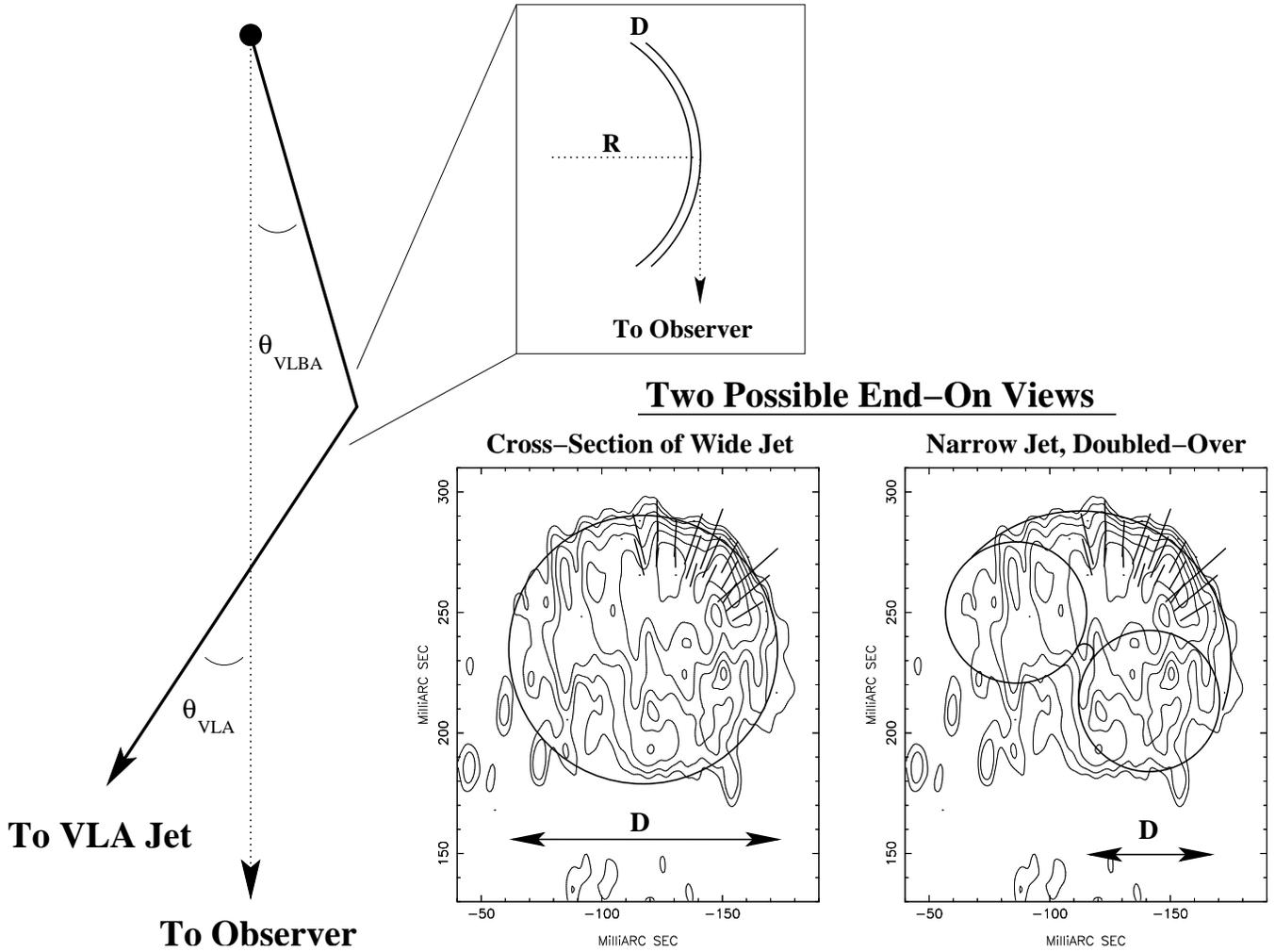,width=7in,angle=0}\\
\end{center}
\figcaption[f3.eps]{\label{f:explain}
Cartoon model for the bent jet geometry.  $D$ is the diameter
of the jet, and $R$ is the radius of curvature of the bend.  Two
possible ``end-on'' views at the point of the bend are depicted
superimposed on our naturally weighted image of the $0.3\arcsec$ 
feature with polarization vectors over-plotted.  The
first is the cross-section of a wide jet pointed directly along our 
line of sight.  The second is of a narrower jet that is not quite
directly along our line of sight, so that it appears to double-over
at the bend.  In this second case, the plane of the jet, as defined by
its two ``straight'' pieces on either side of the bend, does not quite
contain the line of sight.  
}
\end{figure*}

In the second model, where the $0.3\arcsec$ feature marks a 
shocked bend in the jet, the jet is assumed to cross our line of
sight at this point and continues (unseen in our image due to low
surface brightness) to form the arcsecond scale jet.  Figure
\ref{f:explain} depicts a cartoon of this model.  The jet can be
either (1) a wide jet which crosses our line of sight directly so
that we see the cylindrical cross-section of the jet, or (2) a
somewhat narrower jet which crosses our line of sight so that the
plane formed by the bend does not quite contain the line of sight.  
In the second case, the $0.3\arcsec$ feature is formed by the 
doubled over projection of the bent jet, with the jet entering
the bend at the southwestern corner and departing on a parallel 
(but reversed) track from the northeastern corner.  
The high levels of polarization can
be directly due to compression from a shock at the bend
and/or due to shearing of the magnetic field around the bend.  For
shear to have a significant contribution, our line of sight must
be out of the plane of the bend (case 2, above) by at least a
small amount.

\subsection{A Bent Jet}

We favor the `simple' bent jet explanation which ties together the
milli-arcsecond jet, the shocked feature at $0.3\arcsec$, and the
arcsecond scale jet.  Here we explore this model in more detail.  

\subsubsection{The Magnetic Field Order}

\begin{figure*}
\figurenum{4}
\begin{center}
\epsfig{file=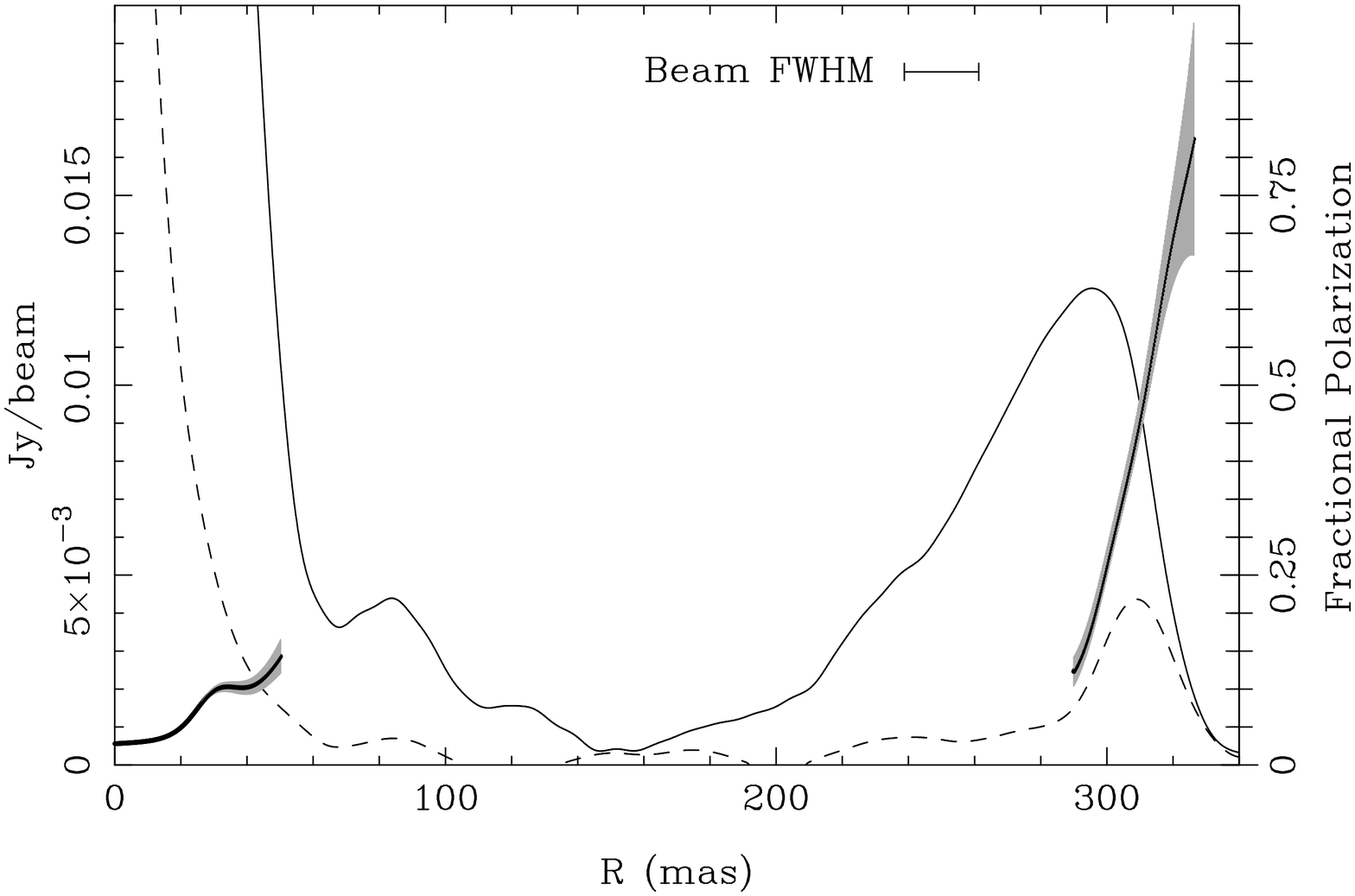,width=6in,angle=0}\\
\end{center}
\figcaption[f4.eps]{\label{f:1510-slice}
Intensity slice through the tapered image of PKS 1510$-$089, 
taken at a position angle of $-28^\circ$ from the VLBI core through
the apex of the feature at $0.3\arcsec$.   Total intensity
is represented by a solid line, polarized intensity is represented
by a dashed line, and fractional polarization is a solid line, 
with a light gray envelope indicating the $2\sigma$ confidence 
interval.  The fractional polarization scale
is given on the right axis of the plot.
}
\end{figure*}

If the magnetic field order is produced entirely by a shock at
the bend, then to generate the observed $\gtrsim 65$\% linear
polarization, the field must be almost perfectly ordered at
that point which requires a {\em very} strong shock \citep*[e.g.][]{HAA85}.  
Strongly shocked emission is greatly enhanced in intensity, and this is
consistent with the high surface brightness of the feature; however, 
the profiles of the intensity and polarization seem inconsistent
with a simple shock model. Figure \ref{f:1510-slice} shows a one
dimensional slice through our tapered 1.7 GHz image taken along a 
position angle of $-28^\circ$ from the VLBI core through the apex of 
the feature at $300+$ milli-arcseconds.  The polarization
occurs at the very outside of the feature; however, the intensity 
peaks further back from the outside edge and falls off relatively 
slowly across the width of the feature.  

In a simple compression
shock, the intensity and polarization enhancement would be expected
to coincide \citep[e.g.][]{HAA85}.  Here the polarization enhancement  
(as seen from our viewing angle) appears confined to the very outside of
the bend.  It seems possible that aberration in a highly relativistic
flow could distort viewing angles enough before and after the bend
to confine the high polarization to the very outside edge where the
range of viewing angles is the narrowest; however, simple models we
created with streaming flow around a semi-circular bend do not have the
polarization falling off fast enough from the outside edge.  These 
simple models considered only the wide jet case, where the 
plane of the bend includes the line of sight.  More complex bends,
where the line of sight does not include the plane of the bend will
have greater possibilities (including a wider range of superposition
effects) for reducing the projected field order away from the outside
of the bend.  

If the field order is produced entirely by a very strong shock, we
must be viewing the shock plane nearly edge-on to see $\gtrsim
65$\% polarization.  For a localized shock, this requires another 
fortunate coincidence of viewing angles in a source which is 
already unusual.  If the shock is spread out along the bend to 
some degree, the necessary coincidence in viewing angles is lessened; 
however, it is difficult to imagine a very strong shock being spread
out in such a fashion.

If the jet is somewhat narrower than the $0.3\arcsec$ feature and the
line of sight does not include the plane of the bend, then at
least some (and perhaps most) of the magnetic field order can be 
produced through shear.  Shear will stretch the magnetic field lines
along the jet.  At the apex of the bend (as viewed in projection)
the local jet direction is perpendicular to that in the
milli-arcsecond jet; therefore, any sheared field would also 
appear perpendicular to the milli-arcsecond jet.  The sheared field
should appear to curve around the apex of the bend, and this
is precisely what we observe.  One concern is that we might expect to 
see polarization along more of the feature as shear would be occurring 
along the entire length of the bend; however, it is only near the apex 
of the projected bend that we see the magnetic field order
undiluted by superpositions with other parts of the jet.  
We note that even with shear producing (at least some
of) the observed magnetic field order, we still need shocked emission 
in the bend to explain the intensity enhancement and spectral index.

\subsubsection{The Bending Angle}

While forming a bend nearly $180^\circ$ in projection, the 
intrinsic bending angle could be quite small.  The close alignment 
of the jet with the line of sight magnifies not only intrinsic bends,
but also the apparent opening angle of the jet, 
$\phi_{obs} = \phi/\sin\theta$ where $\theta$ is the angle the
jet makes with the line of sight.  As shown in section \ref{s:milli}, 
before the bend the intrinsic full-opening angle of the jet was 
$\phi \lesssim 1^\circ$.  The jet's intrinsic opening angle may 
change in the bend, perhaps flaring outward due to a reduction in 
Mach number \citep{ML02}, and we parameterize the intrinsic opening 
angle after the bend to be $\lesssim 1\eta$ degrees.  The apparent 
opening angle after the bend is $\phi_{obs} \gtrsim 10^\circ$ 
from the size of the knot at $2.6\arcsec$.  Taken together, the
intrinsic and observed opening angles give a jet angle to the line of 
sight after the bend of $\theta_{VLA} = \phi/\phi_{obs} 
\lesssim 6\eta$ degrees.

We can get an independent constraint on $\theta_{VLA}$ from the
de-projected size of the $9\arcsec$ jet, which in projection 
appears just 45 kpc long.
In their study of Doppler beaming in large scale jets, \citet{WA97} 
place a rough upper limit on the size distributions of quasar jets 
corresponding to a single-sided jet length of 420 kpc if the 
parent population for quasars cuts off at a viewing angle limit
of $60^\circ$.  In a similar study, \citet{HAP99} find that a normal
distribution fits their data well with a central value corresponding
to a single-sided jet length of $210$ kpc and a standard deviation
of $85$ kpc.\footnote{The
values from both the \citet{WA97} and \citet{HAP99} papers have been
converted to our choice of cosmology.}  Taken together, these values
require $\theta_{VLA} > 6^\circ$ with a reasonable range between
$9^\circ$ and $21^\circ$ to the line of sight.  Given that the 
milli-arcsecond jet is at an angle of $3^\circ$, we estimate 
the intrinsic bending angle is between $12^\circ$ and $24^\circ$.

\subsubsection{What Caused the Bend?}

As noted in \S{\ref{s:milli}}, the bend lies at a de-projected
distance of 30 kpc from the central engine.  Optically,
PKS 1510$-$089 is dominated by a strong point source 
\citep{HCC84}, and the size 
of its host galaxy has not yet been determined \citep{KFS98}.  Given
the typical scale size of the elliptical hosts of radio galaxies, 
$\simeq 30$ kpc \citep{TDH96}, the bend likely lies near the 
boundary of the host galaxy.  The jet could be bent by an 
interaction with a massive cloud, ram pressure 
from winds in the intracluster medium, or simply the density gradient
between the interstellar and intergalactic medium if the gradient
is non-uniform or if the jet crosses it at an oblique angle
(P. Hughes, private communication).
Discriminating between these possibilities 
is difficult; however, whatever the cause, the bend must be stable
on the $\sim10^7$ yr timescale necessary to form the southern 
arcsecond jet.  

Hydrodynamical simulations show that jet-cloud interactions usually
end with the disruption of either the cloud
or the jet \citep[e.g.][]{D91}, and only relatively dense, slow jets
are capable of producing fairly long lived bends via jet-cloud
collisions \citep{WWH00}.  \citet{W02} present evidence that 
PKS 1510$-$089 has a light jet with a particle population dominated
by electron-positron pairs. Combined with the highly relativistic nature
of the milli-arcsecond jet and the location of the 
bend near the probable boundary of the host galaxy, we consider it more
likely that the jet is bent after it leaves the galaxy, either 
by ram pressure from winds in the intracluster medium or by 
the density gradient in the transition to the intergalactic medium.  

\section{Conclusions}

We find that the apparent $\simeq 180^\circ$ misalignment between 
the milli-arcsecond and arcsecond radio jets in PKS 1510$-$089 is most
likely due to a small angle intrinsic bend of
$\simeq12^\circ-24^\circ$, viewed in 
projection.  Our new images show the highly superluminal VLBI
jet to point directly at and nearly connect with a previously
detected ``counter-feature'' to the arcsecond jet, leaving little
doubt that this feature at $0.3\arcsec$ is actually
part of the approaching jet.  We imaged the feature in detail and 
find it to be dominated by shock enhanced emission and to have 
very high levels of polarization confined along its outside edge.  
The projected magnetic field must be almost perfectly ordered at 
this point, ruling out a simple bow-shock model where the $0.3\arcsec$
feature marks the terminal spot in the VLBI jet.  Instead, we prefer 
a shocked-bend model where the jet bends across our line of sight
at this point, creating highly ordered magnetic field either as
a direct consequence of the shock itself or through shear.  The 
bend occurs at a de-projected distance of $\simeq 30$ kiloparsecs 
from the nucleus, near the probable boundary of the host galaxy.
The cause of the bend
is uncertain; however, we favor a scenario where the jet is bent
after it departs the galaxy, either by ram pressure from winds
in the intracluster medium or by the density gradient in the 
transition to the intergalactic medium.  

\section{Acknowledgments}

The authors would like to thank H. Aller, P. Hughes, and M. Lister
for helpful comments and discussion.
This work has been supported by the National Radio Astronomy
Observatory and by NSF Grants AST 95-29228, AST 98-02708 and 
AST 99-00723. This research has made use of
the NASA/IPAC Extragalactic Database (NED) which 
is operated by the Jet Propulsion Laboratory, California Institute of 
Technology, under contract with the National Aeronautics and Space 
Administration. This research has also made use of NASA's Astrophysics 
Data System Abstract Service.



\end{document}